\documentclass[amsmath,onecolumn,superscriptaddress,aps,preprint]{revtex4} \usepackage{amsthm,amsfonts,graphicx,verbatim} \usepackage{graphicx}
\usepackage{bm}

\newcommand{\be}{\begin{equation}} \newcommand{\ee}{\end{equation}} \newcommand{\bea}{\begin{eqnarray}} \newcommand{\eea}{\end{eqnarray}}

  \renewcommand{\epsilon}{\varepsilon} 
   
\renewcommand{\cite}[1]{[\onlinecite{#1}]}  

\begin{document}

\title{Superconductivity at the onset of spin-density-wave order in a metal}
 \author{Yuxuan Wang and Andrey V. Chubukov}
\affiliation{Department of Physics, University of Wisconsin-Madison, Madison, WI 53706, USA}

\begin{abstract} 

We revisit the issue of superconductivity at the quantum-critical point (QCP) between a 2D paramagnet and a spin-density-wave
metal with ordering momentum $(\pi,\pi)$. This problem is highly non-trivial because the system at criticality displays a non-Fermi liquid
behavior and because the effective coupling constant $\lambda$ for the pairing  is  generally of order one, even when the actual interaction is
  smaller than fermionic bandwidth.
Previous study [M. A. Metlitski, S. Sachdev, Phys.Rev.B 82, 075128 (2010)] has found that the 
  renormalizations of the pairing vertex
are stronger than in BCS theory  and hold in powers of 
 $\log^2 (1/T)$, like in color superconductivity. We analyze the full gap equation and argue that, for QCP problem, summing up of the leading logarithms
   does not lead to a pairing instability. Yet, we show that superconductivity  has no threshold 
   and appears
   even if $\lambda$ is
  set to be small, because subleading logarithmical renormalizations diverge
   and give rise to BCS-like  $\log(1/T_c) \propto 1/\lambda$.
 We
 argue that
 the analogy with BCS is not accidental as
 at small $\lambda$
  superconductivity at a QCP predominantly
  comes from fermions which
 retain Fermi liquid behavior at criticality.
 We compute $T_c$ for the actual
 $\lambda \sim O(1)$,
  and found that  both Fermi-liquid and non-Fermi liquid fermions contribute to the pairing.
  The value of  $T_c$ agrees well
  with the numerical results.

 \end{abstract}
  \maketitle

\newpage
{\bf Introduction.} Superconductivity at the onset of density-wave order in a metal is an issue of high current interest, with examples
ranging from
 cuprates~\cite{cuprates}, to
Fe-pnictides~\cite{matsuda}
 and other correlated materials~\cite{cps,chi,rmp}
   It is widely believed that the pairing in these systems
    is
 caused by repulsive electron-electron interaction, enhanced in a
 particular spin or charge channel, which becomes critical at the quantum-critical point (QCP).
 The pairing problem at QCP is highly non-trivial in $D\leq 3$,
  as scattering by a critical collective mode destroys Fermi liquid (FL) behavior above $T_c$ (Ref. \cite{nfl,acs}).
  This is particularly relevant for systems near uniform density-wave instability (e.g., a ferromagnetic or a nematic one).
  In this case, FL behavior is lost on the whole Fermi-surface (FS),  and superconductivity  can be viewed
  as a pairing of incoherent fermions which exchange quanta of gapless collective bosons~\cite{msv,acf,acs,moon,moon-sachdev}.
 The pairing of incoherent fermions is qualitatively different  from BCS/Eliashberg pairing of
 coherent fermions in a FL because
  in the incoherent case the
   pairing in $D <3$
   occurs only if the interaction
   exceeds a certain threshold~\cite{acf,nick,aace}. For $D=3$ there is no threshold, but at small  coupling constant $\lambda$,
   $\log{\Lambda/T_c} =
   1/\sqrt{\lambda}$ rather than  $1/\lambda$ (Ref. \cite{joerg}),
  in close analogy to
  $T_c$ in color superconductivity (CSC) of quarks
  mediated by the exchange of gluons~\cite{color}

 The non-FL behavior at criticality is less pronounced for systems near density-wave order at a finite momentum, because only fermions near
 particular
  points along the FS (hot spots) lose FL behavior at criticality.   Still, fermions from hot regions mostly contribute to the pairing, and
  early studies
   of superconductivity at the onset of $(\pi,\pi)$ spin-density-wave (SDW) order~\cite{acf,acs} placed the pairing problem into the same
   universality class as
   for QCP with $q=0$. The 2D problem has been recently re-analyzed~\cite{ms} by Metlitski and Sachdev (MS). They argued that it is important
  to include into the
   consideration the momentum dependence of the self-energy along the FS, neglected in earlier studies.  Using the full form of $\Sigma
   (\omega_m, {\bf k})$ for
   ${\bf k}$ on the FS, they
   found that the one-loop renormalization of the pairing vertex is larger than previously thought -- it is $\log^2$ instead of $\log$, and
   that the enhancement
   comes from fermions somewhat away from hot spots, for which $\Sigma (\omega_m, {\bf k})$ has a FL form at the smallest frequencies.
   The $\log^2$ behavior in the perturbation theory holds for CSC, and  MS result raises the question whether the pairing problem at a 2D
   SDW QCP is in the same
   universality class as CSC. The related issues raised by MS work are: (i) is the problem analogous to the pairing at a 2D SDW QCP a FL
   phenomenon, or non-FL
   physics is essential, (ii) what sets the scale
   of $T_c$, and (iii) is $T_c$ non-zero only if the coupling $\lambda$ exceed a finite threshold,
   as it happens if one approximates $\Sigma (\omega_m, {\bf k})$ by $\Sigma (\omega_m)$
   at a hot spot, or $T_c$ is non-zero even at smallest $\lambda$, like in CSC?

   In this letter, we address these issues. We first show that the analogy with CSC does not extend beyond one-loop order, and in our case
    the summation of $\log^2$ terms in the Cooper channel does not give rise to a pairing instability.  However,
     that subleading $\log$ terms do give rise to a pairing instability, and
     at weak coupling yield
     $\log \Lambda/T_c 
      \propto1/\lambda$, like in BCS theory. We show that the analogy with BCS formula is
      not accidental
       because the pairing
        at small $\lambda$
       predominantly comes from fermions for which fermionic self-energy has a FL form.
         We then analyze the
       physical case $\lambda = O(1)$
        and argue that in this case
            fermions from both FL and non-FL regimes contribute to the pairing
            and that
            $T_c \approx 0.04 \omega_0$, where $\omega_0$ is the  frequency
            at which $\Sigma (\omega_m)$ at a hot spot becomes equal to $\omega_m$.
         The numerical prefactor agrees with the slope of $T_c$
        obtained by solving the gap equation numerically along the full FS~\cite{acn}.

  {\bf The model.}~~~~
   We follow earlier works\cite{acf,acs,ms} and analyze the pairing near an antiferromagnetic QCP within
  the semi-phenomenological spin-fermion model. The model assumes that antiferromagnetic correlations develop already at
   high energies, of order bandwidth, and mediate interactions between low-energy fermions.
  The static part of the spin-fluctuation propagator is treated as a phenomenological input from high-energy physics, but the
   the dynamical Landau damping part is self-consistently obtained within the model as it
     comes entirely from low-energy fermions~\cite{acf,acs,ms}.
      In the Supplementary material we review  justifications for the spin-fermion model
     and compare spin-fermion approach with the RG-based approaches~\cite{rg,rg_1,rg_2,rg_3} which treat superconductivity,
         magnetism, and specific charge density-wave orders  on equal footings.

We assume, like in~\cite{acf,acs,ms},  that fermions have
   $N \gg 1$ flavors
    and that collective spin excitations are peaked at ${\bf Q} = (\pi,\pi)$, and focus on
    the hot regions  on the FS, i.e., on momenta near ${\bf k}_F$, for which ${\bf k}_F + {\bf Q}$ is also near the FS.
     The
    Lagrangian of the model is
    given by~\cite{cps,acs,ms}
 \begin{eqnarray}
{\mathcal{S}} &=&-\int_{k}^{\Lambda }G_{0}^{-1}\left( k\right) \psi _{k,\alpha }^{\dagger }\psi _{k,\alpha }+\frac{1}{2}\int_{q}^{\Lambda
}\chi _{0}^{-1}\left( q\right) \ {\bf{S}}_{q}\cdot {\bf{S}}_{-q}  \nonumber \\ &&+g\int_{k,q}^{\Lambda }\psi _{k+q,\alpha }^{\dagger }\sigma
_{\alpha \beta }\psi _{k,\beta }\cdot {\bf{S}}_{-q}.\
  \label{startac}
\end{eqnarray}
  where $\int_k^\Lambda$ stands for the integral over $d-$dimensional
 ${\bf{k}}$ (up to some upper cutoff $\Lambda$) and the sum over fermionic and bosonic
  Matsubara frequencies, $G_{0}\left( k\right) = G_0 (\omega_m, {\bf k}) = 1/[i\omega_m - {\bf v}_{F,{\bf k}}  ({\bf k}-{\bf k}_F)]$
  is the bare  fermion propagator, and
   $\chi _{0}\left( q\right) = \chi_0 (\Omega_m, {\bf q}) = \chi_0/({\bf q}^2 + \xi^{-2})$
  is the static propagator of collective bosons, in which $\xi^{-1}$ measures a distance to a QCP and ${\bf q}$ is measured  with respect to
  ${\bf Q}$. We set
  $\xi^{-1} =0$ below.
   The fermion-boson coupling $g$ and $\chi_0$ appear in theory only in combination ${\bar g} = g^2 \chi_0$ and
   we will use ${\bar g}$ below.  The Fermi velocities at hot spots separated by ${\bf Q}$ can be expressed as ${\bf v}_{F,1} = (v_x, v_y)$
   and
   ${\bf v}_{F,2} = (-v_x, v_y)$, where $x$ axis is along ${\bf Q}$. We will also use $\alpha = v_y/v_x$ and
   $v_F = (v^2_x + v^2_y)^{1/2}$.
   The model of Eq. (\ref{startac})
   can be equivalently viewed as a four-patch model for fermions near hot spots at $\pm {\bf k}_F$ and $\pm ({\bf k}_F + {\bf Q})$ (Ref.
   \cite{ms,ms_1}).
   The hot spot model is obviously justified only when
   the interaction ${\bar g}$ is smaller than $E_F$.

The fermion-boson coupling gives rise to fermionic and bosonic self-energies. In the normal state, bosonic self-energy accounts for Landau
damping of spin excitations, while fermionic self-energy accounts for the mass renormalization
  and a finite lifetime of a fermion.  At one-loop level, self-consistent normal-state analysis yields~\cite{acs,ms,millis}
  \bea
 &&\chi (\Omega_m, {\bf q}) = \frac{\chi_0}{{\bf q}^2 + |\Omega_m| \gamma}  \label{2}\\
 &&\Sigma (\omega_m, k_\parallel) = \frac{3 {\bar g}}{4 \pi v_F}~\frac{2 \omega_m}{\sqrt{\gamma |\omega_m| +
 \left(\frac{2 k_\parallel \alpha}{1+
 \alpha^2}\right)^2} + \left|\frac{2 k _\parallel \alpha}{1 + \alpha^2}\right|},
  \label{2a}
 \eea
  where $\gamma = 2N {\bar g}/(\pi v_x v_y)$ and $k_\parallel$ is a deviation from a hot spot along the FS.
   The bosonic propagator $\chi (\Omega_m, {\bf q})$ describes Landau-overdamped spin fluctuations.
  The fermionic self-energy has a non-FL form right at a hot spot:
    $\Sigma (\omega_m,0) =
   (|\omega_m| \omega_0)^{1/2} {\rm sgn} \omega_m$,
   where $\omega_0 = (9 {\bar g}/(16 \pi N)) (2 v_x v_y/v^2_F)$. Away from a hot spot, $\Sigma (\omega_m, k_\parallel)$ retains a FL form at the smallest $\omega_m$ and scales as
    $\Sigma (\omega_m, k_\parallel) \propto \omega_m/|k_\parallel|$.

We use Eqs. \ref{2} and \ref{2a} as inputs for the pairing problem and neglect higher order terms in the loop expansion.
Most of higher-order terms are small in $1/N$, but some terms with $n \geq 4$ loops do not contain $1/N$ (Refs. \cite{ms,sslee,ms_1}).
The terms without $1/N$ include, in particular, feedback effects from
 pairing fluctuations on the fermionic and bosonic propagators.  We verified that these feedback effects preserve the forms of $\chi$ and $\Sigma$, and we just assume  that they do not substantially modify the prefactors.

  {\bf The pairing vertex}~~~~ We add to the action the anomalous term
  $\Phi_0 (k) \psi _{k, \alpha } (i\sigma^{y})_{\alpha \beta
}\psi _{-k,\beta }$ and use Eq. (\ref{startac}) to renormalize it into the full $\Phi (k)$.
 At $T_c$, the pairing susceptibility $\chi_{pp} (k) = \Phi (k)/\Phi_0$ must diverge for all $k$.  The bare $\Phi_0$ can be set constant
 within a patch, but has
 to change signs between patches separated by ${\bf Q}$ (the pairing symmetry at the onset of SDW order is a $d-$wave~\cite{Scal}).
 The one-loop renormalization of  $\Phi (k)$ at $k=(\omega \sim T, 0)$ was obtained by MS:
 \be
 \Phi (\omega \sim T, 0) = \Phi_0 \left(1 + \frac{\lambda}{2\pi} \log^2{\Lambda/T}\right), ~~\lambda = \frac{2 \alpha}{(1 + \alpha^2)},
 \label{1}
 \ee
where $\Lambda$ is the smaller of $\omega_0$ and $\alpha^4E_F^2/\omega_0$.
 Notice that neither the coupling constant ${\bar g}$ not $1/N$ appear in (\ref{1}), the only parameter is the ratio of the velocities
 $\alpha$, which is a
 geometrical property of the FS. For a cuprate-like FS,  $\alpha \sim 1$, i.e.,
 the pairing coupling constant $\lambda = O(1)$.
 To understand the physics of the pairing
 at the QCP, we find that it is instructive  to  formally replace $\lambda$ by $\epsilon \lambda$  and
 first analyze the pairing in the ``weak coupling" case
  $\epsilon \ll 1$.

\begin{figure}[htbp] \includegraphics[width=0.5\columnwidth]{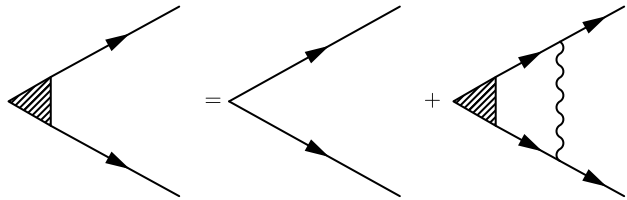} \caption{Diagrammatic representation for the pairing vertex. The shaded
triangle is the full $\Phi_k$, the unshaded vertex is the bare $\Phi_0$, solid lines are full fermionic propagators, and the wavy line is the
Landau-overdamped spin propagator. The pairing vertex contains $i\sigma^y_{\alpha,\beta}$, the vertices where wavy and solid lines meet
contain ${\bf \sigma}_{\gamma\delta}$.} \label{fig:1} \end{figure}
 Let's first
 see
 where $\log^2$ renormalization comes from.  The one-loop diagram for $\Phi$ contains two fermionic propagators
$G(k)$ and $G(-k)$ and one bosonic $\chi (k)$ (Fig.\ref{fig:1}).  Large $N$ allows one to restrict  $\chi (\Omega_m,{\bf k})$ to momenta
connecting points at the FS and integrate over momenta transverse to the FS in the fermionic propagators only. Because $\Sigma$ does not
depend on this momentum, the integration is straightforward, and yields, to logarithmic accuracy $\int GG \chi \propto \int dk_\parallel
\int_T d \Omega_m (\chi (\Omega_m, k_\parallel)/|\Omega_m + \Sigma (\Omega_m, k_\parallel))|$. At $k^2_\parallel > \gamma \Omega_m$ and
$|k_\parallel| < k_F {\bar g}/v_F$,
 $1/|\Omega_m + \Sigma (\Omega_m, k_\parallel)|$ scales as $|k_\parallel/\Omega_m|$ and $ \chi (\Omega_m, k_\parallel) \propto
 1/k^2_\parallel$.
  Integrating over $k_\parallel$ we obtain $\int_{\gamma |\Omega_m|} dk^2_\parallel/k^2_\parallel \propto \log |\Omega_m|$, and
   the remaining integral over
  frequency yields $\int GG \chi \propto \int_T (d \Omega_m/|\Omega_m|)~\log{|\Omega_m|} \propto \log^2 T$.   We see that the $\log^2 T$
  dependence
     originates from extra logarithm from $k-$integration. This
       extra logarithm is in turn the consequence of $\Omega_m/k_\parallel$ form of self-energy $\Sigma (\Omega_m, k_\parallel)$ at
       $k^2_\parallel > \gamma
      \Omega_m$.  As $\Sigma \propto \omega$ is the property of a FL,
        the $\log^2T$ renormalization comes from fermions which preserve a FL behavior at a QCP.
    We further see that the one-loop renormalization can be interpreted as coming from the process in which
    fermions are exchanging quanta of an effective local $\log \Omega$ interaction. The same process determines one-loop renormalization of
    $\Phi$ in CSC.

 The $\log^2$ analysis can be extended beyond leading order. We assume that $\lambda = 2 \epsilon \alpha/(1+ \alpha^2)$ is small
  (because we set $\epsilon$ to be small), but $\lambda \log^2 T = O(1)$, and sum up
  ladder series of $\lambda \log^2 T$
 terms, neglecting smaller
  powers of logarithms
   at each order of loop expansion.
   Performing the calculations (see Supplementary material for details),
    we find that the analogy with CSC does not extend beyond leading order: for CSC the summation of $\lambda \log^2 T$ terms yields
 $\Phi = \Phi_0/\cos[(2\lambda \log^2 T)^{1/2}]$ (Ref.\cite{joerg}), and the system develops a pairing instability at $|\log T_c| =
 \pi/2\sqrt{2\lambda}$
  (Ref. \cite{color}). In our case, perturbation series yield  $\Phi = \Phi_0 e^{\lambda/2\pi \log^2 T}$, i.e., the pairing susceptibility
  increases with decreasing
  $T$, but never diverges.  Because the summation of the leading logarithms does not lead to a finite $T_c$,
  one has to go beyond the leading logarithmical approximation and analyze the full  equation for $\Phi (k)$ at $\Phi_0 =0$ in order to
  understand  whether or
  not $T_c$
   is finite at a QCP. This is what we do next.

\begin{figure}[t] $\begin{array}{cc} \includegraphics[width=.3\columnwidth]{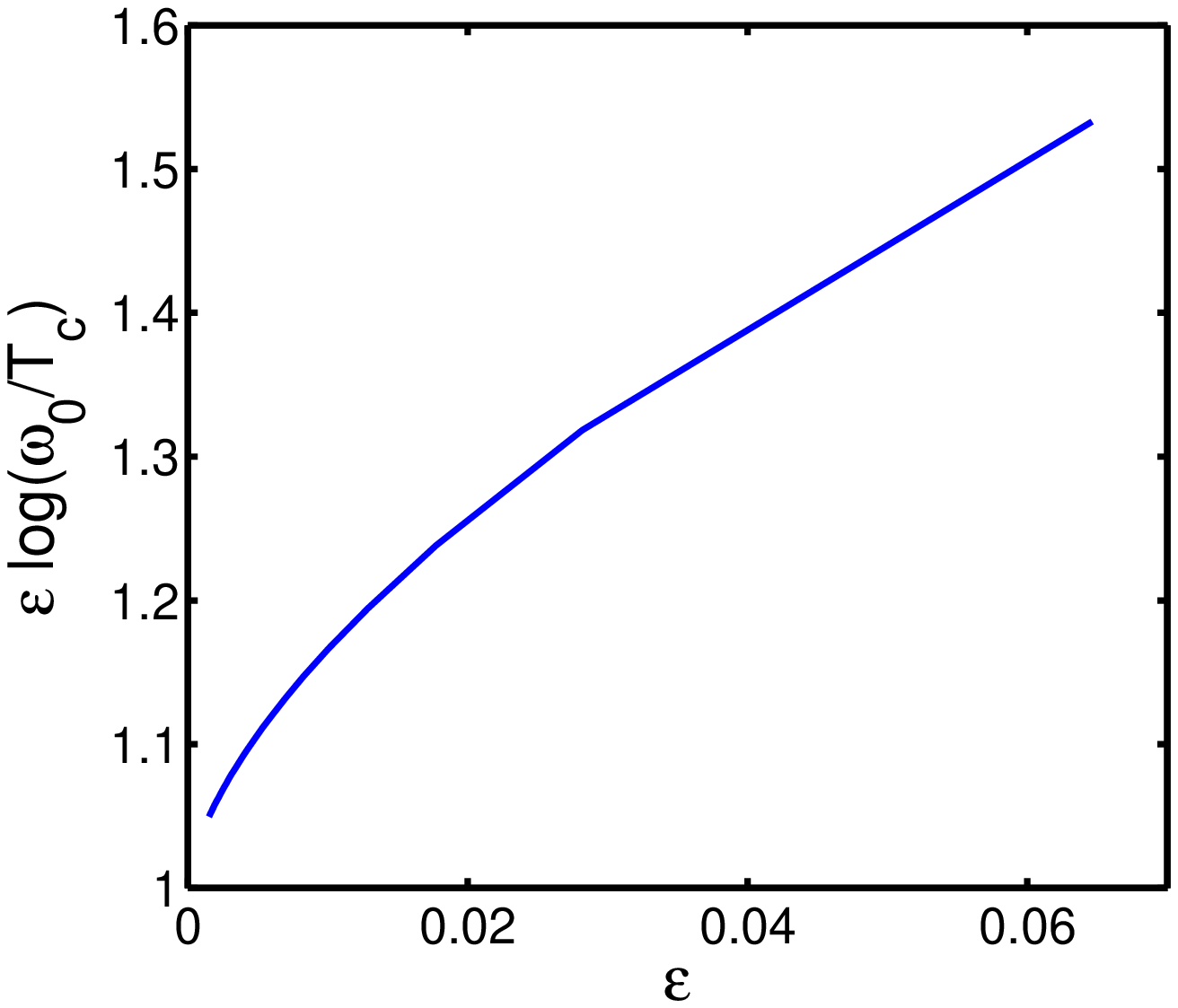}& \includegraphics[width=.3\columnwidth]{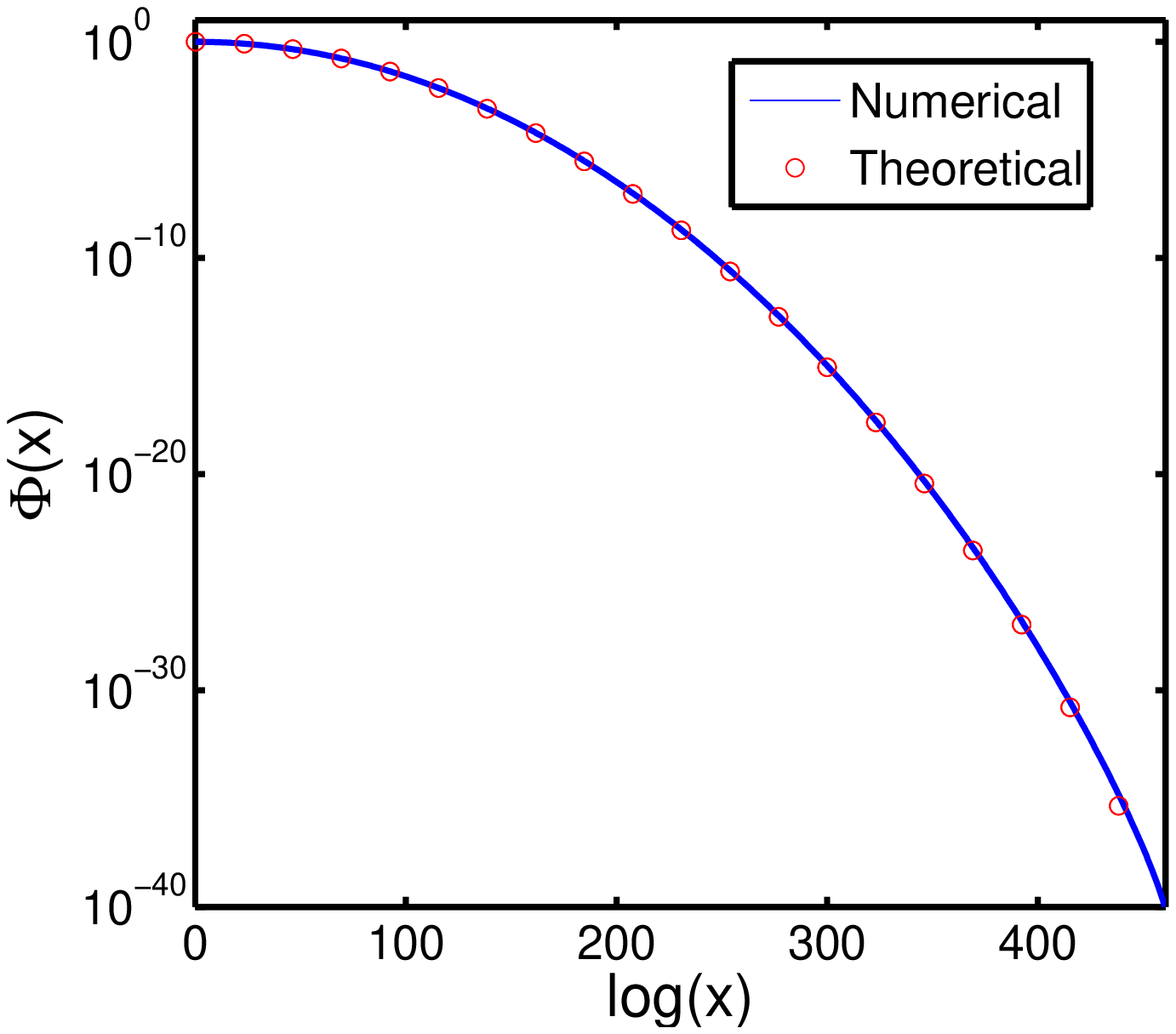}
\end{array}$ \caption{Numerical solution of Eq. (\ref{4}) at small $\epsilon$. (a) The transition temperature. When $\epsilon$ decreases,
$\epsilon \log \omega_0/T_c$ approaches $1$, as in Eq. (\ref{5}).
 (b) The eigenfunction $\Phi (y)$, where $y = k^2_\parallel/(\pi T\gamma)$.
Solid and dashed lines are numerical and analytical solutions of Eq. \ref{4}, respectively.
 The two are very close, except for the largest $y \sim \omega_0/T$, when the cutoff becomes relevant.}
\label{fig:phinum} \end{figure}

  {\bf Full gap equation.}~~~
 Within our approximation, the full linearized equation for the anomalous vertex is obtained by summing up ladder diagrams and keeping the
 self-energy in the
 fermionic propagator.
 Integrating the r.h.s. of this equation  over momenta transverse to the FS, we obtain
 \bea
&&\Phi(\omega_m, k_\parallel)= \frac {3 {\bar g}}{2 v_F} T\sum_{m'}\int\frac{d k'_\parallel}{2\pi} \frac{\Phi(\omega_{m'}, k^{'}_{\parallel})}
{ |\omega_{m'} +\Sigma(\omega_{m'}, k^{'}_{\parallel})|} \nonumber \\
 &&
 \times \frac{1}{k^2_\parallel + k^{'2}_\parallel -2
 \mu k_\parallel k^{'}_{\parallel} +\gamma |\omega_m-\omega_{m^{'}}|}
\label{pairing} \eea where $\mu = (1-\alpha^2)/(1+\alpha^2)$. The temperature at which the solution exists is $T_c$. The overall factor
${3\bar g}/(2v_F)$ is eliminated by rescaling and get replaced by  $\lambda$, which, we recall, we treat as a small parameter.
 One can verify that
 typical $k^2_\parallel$ are larger than typical $\gamma \omega_m$, and
  that
   the vertex $\Phi (\omega_m, k_\parallel)$ has a stronger dependence on $k_\parallel$ than on frequency. In this situation, one can
      approximate $\Phi (\omega_m, k_\parallel)$ by $\Phi (k_\parallel)$, explicitly sum up over frequency and reduce (\ref{pairing}) to 1D
      integral equation.

  For simplicity, we first consider the case when $\alpha=1$, i.e $\lambda = \epsilon$.
  Introducing
  ${\bar T}= \pi T/\omega_0$ and $x = k^2_\parallel/(\gamma \omega_0 {\bar T})$,
   we obtain from (\ref{pairing})
 \begin{align}
\Phi(y)=\frac\epsilon\pi \int_1\frac{dx}{x+y} \frac{\log{x}}{2 \sqrt{x {\bar T}} +1}  \Phi (x) \label{4} \end{align} The term in the
denominator with $\sqrt{x{\bar T}}$ is a soft upper cutoff.

 The r.h.s. of (\ref{4}) contains $\log^2$ contributions from the range $x \gg y$, but, as we
  just found,  they do not lead to a pairing instability. We therefore focus on the
 contribution from $x \sim y$. Because the kernel is logarithmical, we search
  for $\Phi (x)$ in the form
  $\Phi(x)= \exp[-f(p(x))]$, where $p(x) = \epsilon \log{x}$.
 Substituting this into (\ref{4}), we find that the form is reproduced
  at  $1 \ll x \ll 1/{\bar T}$, when soft cutoff can be omitted.
  The self-consistency condition yields (see Supplementary material)
 \be f(z) = \frac{1}{\pi \epsilon} \left(z \arcsin{z} + \sqrt{1-z^2} -1\right).
 \ee
   At small $\epsilon$, the soft cutoff
  can be replaced by the
  boundary condition that  $df(z)/d z$ must be
  at a  maximum
  at $z = \epsilon |\log {\bar T}|$.
   This  condition
   sets
   \be
   T_c \sim \omega_0 e^{-1/\epsilon}.
  \label{5}
  \ee
  To verify this reasoning, we solved Eq. (\ref{4}) numerically and found very good agreement with analytical results (see
  Fig.\ref{fig:phinum}).

    We next analyze the gap equation at $\alpha\neq 1$
     Using the same logic as before we find (see Supplementary material for details) that
      Eq. (\ref{5}) does not change, i.e.,
      to
     logarithmical accuracy,
     $T_c/\omega_0$ does not depend on the angle between Fermi velocities at ${\bf k}_F$ and ${\bf k} + {\bf Q}$.  We
      verified the independence of $T_c/\omega_0$ on $\alpha$  by solving Eq. (\ref{4}) numerically for different $\alpha$.

  We see
  from (\ref{5})
  that $T_c$ is non-zero already at infinitesimally small coupling, like in BCS theory.
   The analogy is not accidental as the pairing
  predominantly comes from momenta away from hot spots, for which
  $x \sim y \sim {\bar T}$, i.e., $k_{\parallel} \sim k_{\perp} \sim (\gamma \omega_0)^{1/2}$.
      Because $T_c \ll \omega_0$, typical $\gamma\omega \geq \gamma T_c$ are much smaller than $\gamma \omega_0$, hence
       fermionic  self-energy  for $k_\parallel \sim (\gamma \omega_0)^{1/2}$
    has the FL form $\Sigma (\omega_m, k_\parallel)\propto \omega_m/|k_\parallel|$. Furthermore,
     for $x\sim y$ in (\ref{4}), the integration over $x$ does not give rise to an additional logarithm besides $\log x$, which is a Cooper
     logarithm.
     The instability at $T_c$ is then a  conventional Cooper instability of a FL with a weak and non-singular attractive coupling $\epsilon$.
      In other words, for small $\epsilon$, the pairing at a SDW QCP
    is entirely a FL phenomenon.

 Although Eq. (\ref{5}) looks like BCS formula, the problem we are solving is not a weak-coupling pairing by a static attractive interaction. We
 emphasize in this regard that a non-zero $T_c$ at small $\epsilon$ is
 the consequence of the dependence of the self-energy on
 the momenta along the FS.  Earlier works~\cite{acf,acs} neglected this momentum dependence and
  approximated the self-energy by its non-FL form
 $\Sigma (\omega) = \omega_m (\omega_0/|\omega_m|)^{1/2}$ at a hot spot.
  These studies found a different result:
   $T_c$ at an AFM QCP becomes non-zero only if $\epsilon$ exceeds a certain threshold,
    like in  the pairing problem at a QCP with $q=0$ (Refs.\cite{aace,max_last}).
  Specifically, for $\Sigma = \Sigma (\omega_m)$, the anomalous vertex $\Phi$
  also depends only on frequency, and
 Eq. (\ref{pairing}) reduces to 1D integral equation in frequency rather than in momentum:
 \begin{align}
\Phi(\omega_m)=  \frac{\pi\epsilon T}{2} \sum_{m'\neq m} \frac{\Phi(\omega_{m'})}{\sqrt{|\omega_{m'}|} Z_{\omega_{m'}}
\sqrt{|\omega_{m}-\omega_{m'}|}}. \label{VerEqLog1} \end{align} where  $Z_{\omega_{m'}} = 1+ \sqrt{|\omega_{m'}|/\omega_0}$.
 This equation has been solved for
 arbitrary ${\epsilon}$~~\cite{acf}, and the result is that $T_c$ becomes non-zero only when ${\epsilon}$ exceeds a critical value
 ${\epsilon}_c =0.22$.
   Near critical coupling $T_c \sim \omega_0
 e^{-3.41/({\epsilon} - {\epsilon}_c)^{1/2}}$, and for
  ${\epsilon} =1$, $T_c = 0.17 \omega_0$.

{\bf $T_c$ at moderate coupling.}~~~~
 For the actual physical case  $\epsilon =1$
  we solved  Eq. \ref{pairing} numerically and found
 that the behavior of $T_c (\alpha)$ is very similar to that at small $\epsilon$.
Namely, $T_c$ scales with $\omega_0$ and
 the prefactor is essentially independent on $\alpha$ as long as $\alpha \gg {\bar g}/E_F$.
 We obtained
 \be
T_c \approx 0.04 \omega_0. \label{6} \ee
For  $\epsilon =1$,
 typical
 $(\alpha k_\parallel)^2 \sim \gamma \omega_0$ and typical $\gamma \omega \sim \gamma T_c$ are now comparable, i.e., for $\epsilon =1$
the pairing comes from fermions whose self-energy is in a grey area between a FL and a non-FL.
 We checked this by solving for $T_c$ using the two limiting forms of the self-energy in Eq. (\ref{2}) -- the non-Fl $\Sigma (\omega_m)$
 right at a hot spot
(this gives $T_c \sim 0.17 \omega_0$) and the FL form $\Sigma (\omega_m, k_\parallel) \propto \omega_m/k_\parallel$ (this gives $T_c =
 0.005 \omega_0$). The
 actual $T_c$ given by Eq. (\ref{6}) is in between the two limits.
 We also verified (see Supplementary material) that in the extreme case of strong nesting, when $\alpha$ gets smaller than $({\bar g}/E_F)$,
 the momentum dependence of the self-energy becomes irrelevant for all $k_\parallel$ along the FS, and $T_c$ crosses over to
 $T_c \sim 0.17 \omega_0$.

The linearized gap equation has been previously solved numerically along the full FS, without restriction to hot spots~\cite{acn}. In
notations of Ref. ~\cite{acn},
  $T_c = (v_F/a) f(u)$, where dimensionless $u = 4\omega_0 a/(3v_F)$. Eq. (\ref{6}) implies that $f(u) =0.03 u$ at small $u$.
  This  agrees well with the numerical solution in~\cite{acn}. At larger $u \geq 1/2$, $f(u)$  saturates at
   around $0.015-0.02$ (Refs. \onlinecite{acn,numerics}), and at larger $u$ decreases as $1/u$ because of Mott physics.

 {\bf Conclusions.}~~~ In this paper we analyzed the equation for superconducting $T_c$ at the
  onset of SDW order in a 2D metal. We demonstrated that the
   leading perturbation correction to the bare pairing vertex contains $\log^2 T$, but
   the series of
   $\log^2 T$ renormalizations
    do not give rise to the pairing instability.
    Yet, $T_c$ is finite, even when coupling $\lambda$ is artificially set to be small, because of
 subleading, $\log T$ terms. We showed that for physical $\lambda = O(1)$,
  the pairing at a QCP comes from fermions
  with both FL and non-FL forms of the self-energy.
  The overall scale of $T_c$ is set by the
  interaction ($\omega_0 \sim {\bar g}$), as long as the interaction is smaller than the Fermi energy, and the prefactor is
  essentially independent on the details of the geometry of the FS.

The issue which requires a further study is how robust these results are with respect to feedback effects from  pairing fluctuations on
the fermionic and bosonic propagators.  These feedbacks are quite relevant in the RG-based studies~\cite{rg,rg_1,rg_2,rg_3}.
  In the spin-fermion model, the corrections from the pairing channel come from diagrams with $n\ge 4$ loops and are not small in $1/N$. These corrections preserve the Landau-overdamped form of the bosonic propagator and the $\omega/k_\parallel$ form of the fermionic self-energy, but may contribute additional logarithm $\log{k^2_\parallel/(\gamma |\omega|)}$ to $\Sigma$ (see Ref.~\cite{ms_1} and Supplementary material).  The argument of the logarithm is, however, of order one for typical $k_\parallel$ and $\omega$ in the calculations of $T_c$,  hence we expect that the feedbacks from the
 pairing channel will at most change the prefactor for $T_c$ but do not change our two main conclusions that (i) $T_c$ scales with $\omega_0$, and (ii) in the physical case the pairing involves fermions with both FL and non-FL forms of the self-energy. It is very likely that the same conclusions can be reached within RG-based approaches as the results of the RG analysis are generally comparable to those obtained in the spin-fermion model~\cite{rg_3}.

 We acknowledge stimulating discussions with Y.B. Kim, S.S. Lee, M.A. Metlitski, S. Sachdev, T. Senthil, and A-M Tremblay. The research has
been supported by DOE  DE-FG02-ER46900.

\newpage

\section{Supplementary Material}

\section{Spin-fermion model and its relation to RG-based theories}

For a generic metal with no nesting of the Fermi surface, an instability towards a magnetic order does not occur at weak coupling, but may
occur when the interaction $U$ becomes of order of fermionic bandwidth, $W$.
In this intermediate coupling regime, exact analytical treatment is hardly possible and one has to rely on approximate computational schemes
 in a hope that an approximate treatment still captures the key physics of the underlying microscopic model.
 The spin-fermion model is a ``minimal" semi-phenomenological effective low-energy model of this kind.
  It takes as inputs the experimental Fermi surface and
 the experimental fact that the underlying system does have a transition between a metallic paramagnet and a metallic antiferromagnet (often called a spin-density-wave (SDW)).  Since Fermi surface shows no nesting, the underlying interaction must be of order
  bandwidth, which in turn implies that  antiferromagnetic correlations come from  fermions with energies comparable to the bandwidth.
 The spin-fermion model is applicable if antiferromagnetism is the {\it only} instability which develops already at energies comparable to a bandwidth. There may be other instabilities (e.g., superconductivity or charge order), but the assumption is that they develop at energies smaller than the upper end set for the spin-fermion model and can be fully understood within the low-energy theory.

 The Lagrangian of the spin-fermion model (Eq. (1) in the main text) contains three terms describing fermions, collective spin excitations, and the minimal coupling between a spin of a fermion and a spin of a collective excitation.  Such a Lagrangian can be formally derived within RPA, starting from a Hubbard model~\cite{cm1}, but this only serves an illustration purpose because RPA is an uncontrolled approximation for $U \sim W$.
 A more sophisticated justification of the spin-fermion model with respect to, e.g., the Hubdard model near optimal doping, comes from numerical studies~\cite{scalapino1} in which the pairing interaction and the dynamical spin susceptibility were calculated independently and, to a good accuracy, were found to be proportional to each other.

 The description within the low-energy spin-fermion model makes sense if the interaction does not mix low-energy and high-energy sectors. This is true if the residual interaction between low-energy fermions (${\bar g}$ in the notations of the main text) is smaller than the upper energy cutoff of the theory, which is generally a fraction of the bandwidth.  At a face value, such an approximation is not consistent
  with the initial assumption that the underlying, microscopic interaction $U$ is of order bandwidth.  There are several ways to make the assumptions
   ${\bar g} \ll W$ and $U \sim W$  consistent~\cite{cm1} (e.g., if microscopic interaction has a length $a$ which is large enough such that $ak_F \gg 1$, ${\bar g}$ is small in $1/(ak_F)$ compared to $U$).  But since the properties of spin-fermion model do not depend in any singular way on ${\bar g}/W$ ratio,  the hope is that, even for a short-range interaction, spin-fermion model captures the essential physics of the system behavior near an antiferromagnetic instability in a metal, including a pre-emptive superconducting instability.

Several authors~\cite{rg1,rg_11,rg_21,rg_31} put forward another approach, which employs the renormalization group (RG) technique.
  This approach assumes that interactions in all channels, including the SDW one, are small at energies of order bandwidth, but
 evolve as one progressively integrates out high-energy degrees of freedom. This approach departs directly from the underlying microscopic model and treats superconductivity, SDW, and specific charge-density-wave instabilities on equal footings. From this perspective, it is more microscopic and less biased than spin-fermion model. At the same time, the RG approach has its own limitation because it
   assumes that that the tendency towards a SDW order can be detected already at a weak coupling. This is true if the Fermi surface is nested and the bare SDW susceptibility at an antiferromagnetic momentum is logarithmically enhanced.
    Then one can rigorously separate logarithmical and non-logarithmical vertex renormalizations, neglect non-logarithmical ones, derive the set of coupled RG equations for different vertices, and analyze the interplay between different ordering tendencies.  Examples when the SDW susceptibility is
    logarithmically enhanced include cuprates and graphene near van-Hove doping~\cite{rg1}, Fe-pnictides~\cite{rg_11}, and quasi-1D organic conductors~\cite{rg_21}.

\begin{figure}[htbp]
\includegraphics[width=.8\columnwidth]{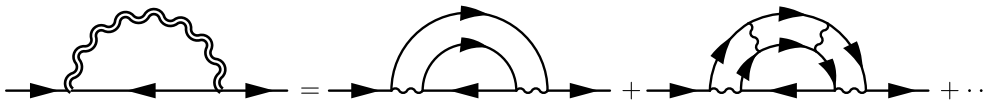}
\caption{The self-energy renormalization from fluctuations in the pairing channel.  The lowest order in the series is a part of
 one-loop self-energy, the new diagrams appear at 4-loop and higher order.  These diagrams are not small in $1/N$ and are of the same order as the one-loop self-energy. These diagrams can only be neglected by numerical reasons.}
\label{fig1_new1}
\end{figure}

\begin{figure}[htbp]
\includegraphics[width=.3\columnwidth]{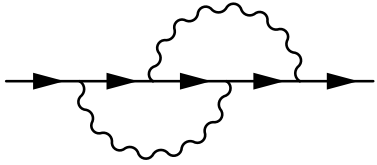}
\caption{The two-loop self-energy diagram with vertex correction. This diagram is small by $1/N$ compared to one-loop diagram. }
\label{fig2_new1}
\end{figure}

     For a generic non-nested Fermi surface, a bare SDW susceptibility
    is, however, not enhanced and generally is of order of inverse bandwidth $1/W$.  Then one needs  large $U \sim W$ to bring the system to the vicinity of an SDW instability, and the terms neglected in the RG scheme become of order one.  Whether to use spin-fermion model or RG technique in this situation becomes a somewhat subjective issue. Our justification of using the spin-fermion model for a 2D metal with cuprate-like Fermi surface is based on a'posteriori argument that superconducting $T_c \approx 0.04 {\omega_0}$ is numerically much smaller than the
     scale $\omega_0$, below which antiferromagnetic correlations substantially modify fermionic self-energy.  The smallness of $T_c/\omega_0$
     then implies that there exists a wide range of energies in which magnetic correlations strongly modify self-energy, yet superconducting fluctuations are still weak.  This was our motivation to compute first fermionic self-energy due to spin fluctuation exchange and then use it
     as an input for the calculation of $T_c$.  If $T_c$ and ${\omega_0}$ were comparable, one could not separate SDW and superconducting channels and had to treat them on equal footing.

\subsection{pairing fluctuations in the spin-fermion model}

A more subtle issue is whether pairing fluctuations can be rigorously neglected in the spin-fermion model, at least
in the limit of large number of fermionic flavors, $N$.  We checked this and argue that the corrections from pairing fluctuations are not small in $1/N$ and can be  neglected only by numerical reasons.  In this respect, the corrections from pairing fluctuations are different from ordinary vertex corrections, as the latter are small in $1/N$.  Our consideration closely follows the one by  Hartnoll et al~\cite{ms_11} who found that  $2k_F$
scattering gives rise to self-energy corrections which are not small in $1/N$.

 The  self-energy due to pairing fluctuations is presented in Fig. \ref{fig1_new1}. The double wavy line represents
  a series of diagrams which contain pairs of fermions with near-opposite frequencies and momenta.  The diagrams  contain even number of bosonic propagators, otherwise one would need to include at least one interaction with a small momentum transfer. The lowest diagram with two propagators is already included into the one-loop self-energy considered in the main text (the extra bubble gives rise to Landau damping term).   At a hot spot, it yields $\Sigma^{(1)} (\omega) \propto ({\bar g} \omega/N)^{1/2}$ (the notations are the same as in the main text). The next in series is the diagram with $n=4$ loops.  It contains four bosonic propagators,  seven fermionic Green's functions, and the integration is over four intermediate 2D momenta and four frequencies.
Fermions in the seven Green's functions are combined into three pairs in the particle-particle channel and one unpaired fermions.  Integrations transverse to the Fermi surface then involve four momentum components. Three integrals involving pairs of fermionic Green's functions yield
 $1/(v_F \Sigma ^{(1)}(\omega'))^3$, where $\omega'$ is one of four internal frequencies, which are all of the same order, the
 fourth integration gives additional $1/v_F$ and restricts
  internal frequencies to be of order on external $\omega$.
Integration over the other four momentum components involves four fermionic propagators and yields $1/(\gamma \omega')^{2}$. The total four-loop self-energy at a hot spot is then
\be
\Sigma^{(4)} (\omega) \propto \left(\frac{\bar g}{v_F}\right)^4 \frac{\omega^4}{(\gamma \omega)^2 (\Sigma^{(1)} (\omega))^3}
\label{ll_11}
\ee
Using $\gamma \sim {\bar g} N/v^2_F$ we find that  $\Sigma^{(4)} (\omega) \propto ({\bar g} \omega/N)^{1/2}$  is of the same order as
$\Sigma^{(1)} (\omega)$, i.e., additional loop order does not give rise to additional powers of $1/N$. One can easily make sure that this holds for all higher order diagrams with $n=6$, $n=8$, etc  loops.

It is instructive to compare this behavior with the effect of a vertex correction. The two-loop self-energy diagram of this kind is shown in
Fig. \ref{fig2_new1}.
It contains three fermionic and two bosonic propagators and integrals over two internal 2D momenta and two frequencies.  The distinction from the previous case is that now one needs three components of momenta to integrate over three fermionic dispersions. These three integrals yield
$1/(v_F)^3$ and restrict internal frequencies to be of order of external $\omega$. The remaining momentum integral involves one bosonic propagator and yields $1/(\gamma \omega)^{1/2}$. This leaves one bosonic propagator and two frequency integrals.  the integration yields $\omega/\gamma$ (up to a logarithm). Combining we find
 \be
\Sigma^{(2)} (\omega) \propto \left(\frac{{\bar g}}{v_F}\right)^2 \frac{\omega}{v_F \gamma (\gamma \omega)^{1/2}}
\label{ll_21}
\ee
 Substituting $\gamma \sim {\bar g} N/v^2_F$, we find that $\Sigma^{(2)} (\omega) \sim \Sigma^{(2)} (\omega)/N$, i.e., ordinary
  vertex corrections are small in $1/N$.

The analysis can be extended to momenta away from a hot spot.  The one-loop self-energy away from a hot spot is $\Sigma^{(1)} (\omega, k_\parallel) \propto \omega/|k_\parallel|$. The self-energy due to pairing fluctuations is of the same form, again without additional $1/N$. There may be extra logarithms in the form $\log {k^2_\parallel/(\gamma \omega)}$ (Ref.\cite{ms_11}), but the argument of the logarithm is $O(1)$ for typical $k_\parallel$ and $\omega$ for the pairing problem (see the main text).

\section{Perturbation theory for the pairing vertex}

\begin{figure}[htbp]
\includegraphics[width=.7\columnwidth]{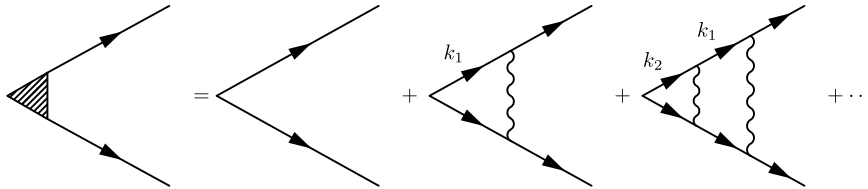}
\caption{Perturbation expansion for the pairing vertex.}
\label{fig:s11}
\end{figure}

We add a fictitious anomalous term $\Phi_0\psi_{k,\alpha}(i\sigma^y)_{\alpha\beta}\psi_{-k,\beta}$ to the original Lagrangian
 to generate a bare pairing vertex, renormalize it by particle-particle interaction, and obtain the pairing susceptibility as a ratio of the
    fully renormalized and bare pairing vertices. Within the approximations which we discuss in the main text, the
     diagrams for the renormalization of the pairing vertex form ladder series, shown in Fig. \ref{fig:s11}, however each line is the full
      fermionic Green's function, which includes the self-energy.

The renormalized vertex $\Phi$ depends on fermionic frequency $\omega_m$, fermionic momentum along the Fermi surface, $k_\parallel$, and on temperature.
 At high enough temperatures $\Phi (\omega_m, k_\parallel, T)$ weakly deviates from $\Phi_0$, but at $T_c$ the ratio $\Phi/\Phi_0$
  should diverge for all frequencies and momenta.  For definiteness, we set $k_\parallel$ to zero and $\omega_m$ to  $\pi T$, i.e., consider
  $\Phi (\pi T, 0, T) = \Phi (T)$.
  To simplify the formulas, we also set the ratio of Fermi velocities $\alpha = v_y/v_x$ to one.

 Consider one-loop renormalization of $\Phi$ (the diagram with one wavy line in Fig. \ref{fig:s11}).
 After integration over momenta transverse to the Fermi surface (FS) we obtain, replacing the summation over $\omega_m$ by integration over
 $|\omega_m| > \pi T$,
 \be
 \Phi (T) = \Phi_0 \left(1+ \pi (\omega_0 \gamma)^{1/2} \int \frac{d\omega_m d k_\parallel}{4\pi^2} \frac{1}{|\omega_m + \Sigma (\omega_m, k_\parallel)|} \frac{1}{k^2_\parallel + \gamma |\omega_m|} \right)
 \label{ch_11}
 \ee
 where, we remind, $\Sigma(\omega_m,k_\|) = {\rm sign} ~\omega_m |\Sigma(\omega_m,k_\|)|$, and~\cite{acs1,ms1}
 \begin{align}
|\Sigma(\omega_m,k_\|)|= \sqrt{\omega_0}\left(\sqrt{|\omega|+ k^2_\parallel/\gamma}
-|k_\parallel|/{\sqrt\gamma} \right).
\label{sigma1}
\end{align}
Rescaling $k_\parallel = (|\omega|\gamma)^{1/2} z$ we obtain
 \be
 \Phi (T) = \Phi_0 \left(1+ \frac{\lambda}{2\pi}  \int^\infty_{\pi T} \frac{d \omega_m}{\omega_m} \int dz \frac{\sqrt{z^2+1} +|z|}{z^2+1}  \frac{1}{1 + \left(\frac{\omega_m}{\omega_0}\right)^{1/2} (\sqrt{1+z^2} +|z|)} \right)
 \label{ch_21}
 \ee
where $\lambda =2\alpha/(\alpha^2+1)=1$.
 One can easily make sure that the the integral in the r.h.s. of (\ref{ch_21})
 contains  $\log^2{\omega_0/T}$, which comes from large $z$, i.e., from $k^2_\parallel \gg |\omega_m| \gamma$ (Ref.\cite{ms1}), and $\log{\omega_0/T}$, which  comes from $z = O(1)$.  The coupling constant $\lambda$ is one,
  so it is not obvious how to go beyond one-loop order.
  We choose a "perturbative" path and artificially make $\lambda$ small by replacing $\lambda$ by ${\bar \lambda} = \epsilon \lambda$ with $\epsilon \ll 1$, such that ${\bar \lambda} \ll 1$.
   Then, obviously, ${\bar \lambda} \log^2{\omega_0/T}$ is more relevant than
 ${\bar \lambda} \log {\omega_0/T}$, i.e., to logarithmic accuracy~\cite{ms1}
 \be
 \Phi (T)=\Phi_0 \left(1+ \frac{{\bar \lambda}}{2\pi} \log^2{\omega_0/T} \right)
 \label{ch_31}
 \ee

 We now check whether the series of ${\bar \lambda} \log^2{\omega_0/T}$ give rise to a pairing instability at
 $\log {\omega_0/T} \sim 1/\sqrt{{\bar \lambda}}$. Because $\log^2{\omega_0/T}$ comes from  $k^2_\parallel \gg |\omega_m| \gamma$, we can expand the self-energy in $\gamma |\omega|/k^2_\parallel$, i.e., approximate $|\Sigma(\omega_m,k_\|)|$ in (\ref{sigma1}) by
 \begin{align}
|\Sigma(\omega_m,k_\|)| \approx 0.5 \sqrt{\omega_0 \gamma} \frac{\omega_m}{|k_\parallel|}
\label{sigma_11}
\end{align}
Using this self-energy, we obtain at two-loop order ($k_{\parallel,1} = k_1, k_{\parallel,2} = k_2$)
\begin{align}
\Phi(T) &=\Phi_0\left(1+
\frac{{\bar \lambda}}{2\pi} \log^2{\omega_0/T} +\frac{{\bar \lambda}^2}{4\pi^2}\int_{\sqrt{\gamma\pi T}}^{\sqrt{\gamma\Lambda}}\frac{2k_1dk_1}{k_1^2}\int_{\pi T}^{k_1^2/\gamma}\frac{d\omega_1}{\omega_1}\int_{\sqrt{\gamma\pi T}}^{\sqrt{\gamma\Lambda}}\frac{2k_2dk_2}{k_2^2+k_1^2}\int_{\pi T}^{k_2^2/\gamma}\frac{d\omega_2}{\omega_2}+O(\lambda^3)\right)
\end{align}
Evaluating the integrals we find
\begin{align}
\Phi(T) &=\Phi_0 \left(1+
\frac{{\bar \lambda}}{2\pi} \log^2{\omega_0/T} +\frac{1}{2} \frac{{\bar \lambda}^2}{4\pi^2}  \log^4{\omega_0/T} + ... \right)
\label{ch_41}
\end{align}
To understand what are the prefactors from higher-order terms, we note that $\log^4$ term comes from the region of internal momenta $k_2 \gg k_1$,
 or, more specifically,  the  momentum $k_1$ in the cross-section which is farther from the vertex sets the lower cutoff of momentum integration over $k_2$
  in the cross-section, which is closer to the vertex.  This sheds light on the general pattern -- we verified that at order $M$ the largest, $\log^{2M} {\omega_0/T}$ term comes from  comes $k_M>k_{M-1}>...>k_1$, where $k_M$ is the momentum in the cross-section, which is the closest to the vertex.
 The same trend was earlier detected in the perturbation theory for the pairing vertex in 2D systems for which fermionic self-energy depends only on frequency~\cite{acf1}.  A simple exercise in combinatorics then yields, for the full series
 \begin{align}
 \Phi(T) &= \left(1+
\frac{{\bar \lambda}}{2\pi} \log^2{\omega_0/T} +\frac{1}{2} \frac{{\bar \lambda}^2}{4\pi^2} \log^4{\omega_0/T} + \frac{1}{6} \frac{{\bar \lambda}^3}{8\pi^3}  \log^6{\omega_0/T} + ... \right) \nonumber \\
&=\Phi_0 \sum^\infty_{M=0} \frac{{\bar \lambda}^M}{(2\pi)^M M!} \log^{2M}\frac{\omega_0}{T} = \Phi_0 e^{({\bar \lambda}/2\pi) \log^2 \omega_0/T}
\end{align}
We see that the calculation to $\log^2$ accuracy does not give rise to a pairing instability: the pairing susceptibility, $\Phi/\Phi_0$ increases with decreasing $T$, but does not diverge at any finite $T$.

\subsection{A comparison with color superconductivity}

The presence of $\log^2 \omega_0/T$ in perturbation theory brings in the comparison with the problem of color superconductivity (CSC),
  where perturbative expansion also contains series of ${\bar \lambda} \log^2 T$ terms. For that problem, however,
  previous works have demonstrated that the summation of $\log^2$ terms does lead to a pairing instability. It is therefore instructive
   to compare perturbation theory for CSC problem with our case and see where the two cases differ.

   In CSC problem~\cite{color1,joerg1}, as well as in condensed-matter problems of the pairing  mediated  by gapless collective excitations in three spatial dimensions~\cite{moon1},
   fermionic self-energy depends only on frequency (and is actually irrelevant to the pairing problem),
    and  the extra logarithm, in addition to a Cooper one, appears because
    the effective interaction, integrated over momenta along the FS, has logarithmic dependence on frequency.
    Because the integration over momenta $k_\parallel$ can be done independently in any cross-section in Fig. \ref{fig:s11}, the pairing can be analyzed within the effective local model of fermions with dynamical interaction~\cite{color1}
    \be
    \chi (\omega_m) = {\bar \lambda} \log{\frac{{\bar \Lambda}}{|\omega_m|}}
    \ee
    Substituting this form of the interaction (wavy line) into the one-loop diagram for $\Phi$, we obtain
 \begin{align}
\Phi(T) =\Phi_0\left(1+ 2{\bar \lambda}\int_{\pi T}^{{\Bar \lambda}}\frac{d\omega}{\omega}\log{\frac{\Lambda}{|\omega|}}\right) = \Phi_0\left(1+ {\bar \lambda}\log^2\frac{\Lambda}{\pi T}\right)
\label{ch_61}
\end{align}
This  is the same result as in our case.
At two-loop order, however, the prefactor for ${\bar \lambda}^2 \log^4{\Lambda/T}$ term is different from the one in our case.
For CSC we have, at two-loop order
\begin{align}
\Phi(T)=&\Phi_0\left(1+{\bar \lambda} \log^2\frac{\Lambda}{\pi T} + {\bar \lambda}^2\int_{\pi T}^{\Lambda}\frac{2d\omega^{'}}{\omega^{'}}\log\frac{\Lambda}{|\omega^{'}|}\int_{\pi T}^{\Lambda}\frac{2d\omega^{''}}{\omega^{''}}\log\frac{\Lambda}{|\omega^{'}-\omega^{''}|}\right)
\end{align}
It is natural to divide the integration range over $\omega^{''}$ into two regimes, $\omega^{''} \gg \omega'$ and $\omega^{''} \ll \omega'$. It turns out that {\it both} regimes give contributions of order $\log^4{\Lambda/T}$.  We have
\begin{align}
&{\bar \lambda}^2\int_{\pi T}^{\Lambda}\frac{2d\omega'}{\omega'}\log\frac{\Lambda}{|\omega^{'}|}\int_{\pi T}^{\Lambda}\frac{2d\omega''}{\omega''}\log\frac{\Lambda}{|\omega^{'}-\omega^{''}|}\nonumber\\
=&{\bar \lambda}^2\int_{\pi T}^{\Lambda}\frac{2d\omega'}{\omega'}\log\frac{\Lambda}{{|\omega^{'}|}}\left(\int_{\pi T}^{\omega'}\frac{2d\omega''}{\omega''}\log\frac{\Lambda}{\omega'}+\int_{\omega'}^{\Lambda}\frac{2d\omega''}{\omega''}\log\frac{\Lambda}{\omega''}\right)\nonumber\\
=&{\bar \lambda}^2\left(2\int_{\pi T}^{\Lambda}\frac{2d\omega'}{\omega'}\log\frac{\Lambda}{{|\omega^{'}|}}\log\frac{\omega'}{\pi T}\log\frac{\Lambda}{\omega'}+\int_{\pi T}^{\Lambda}\frac{2d\omega'}{\omega'}\log\frac{\Lambda}{|\omega'|}\log^2\frac{\Lambda}{\omega'}\right)\nonumber\\
=&{\bar \lambda}^2\left(\frac 1 3\log^4\frac{\Lambda}{\pi T}+\frac 1 2\log^4\frac{\Lambda}{\pi T}\right)\nonumber\\
=&\frac 5 6{\bar \lambda}^2\log^4\frac{\Lambda}{\pi T},
\end{align}
Combining one-loop and two-loop terms, we obtain
 \begin{align}
\Phi(T)&= \Phi_0\left(1+ {\bar \lambda}\log^2\frac{\Lambda}{\pi T} + \frac 5 6{\bar \lambda}^2\log^4\frac{\Lambda}{\pi T}\right)
\label{ch_7_11}
\end{align}
which is different from the two-loop result in our case, Eq. (\ref{ch_41}).  The difference comes about because for CSC there is no
 requirement that highest power of the logarithm comes from particular hierarchy of running frequencies -- at two loop order, there
  is $\log^4$ contribution from the range where the highest frequency in the cross-section next to the vertex, and from the range when the
   highest frequency is in the cross-section farthest from the vertex.

The series of ${\bar \lambda} \log^2$ terms for CSC problem have been summed up in Ref. \cite{joerg1}, an the result is
\begin{align}
\Phi(\omega=\pi T)&=
\Phi_0\left(1+{\bar \lambda}\log^2\frac{\Lambda}{\pi T}+\frac{5}{6}{\bar \lambda}^2\log^4\frac{\Lambda}{\pi T}+O({\bar \lambda}^3)\right)\nonumber\\
&=\frac{\Phi_0}{\cos[(2{\bar \lambda}\log^2\Lambda/\pi T)^{1/2}]}
\label{cscsol1}
\end{align}

\section{Linearized gap equation}

\subsection{Interplay between characteristic momenta and frequency}

We start with the Eq (5) in the main text. We first discuss the case where $\alpha =1$. In this case
${\bar \lambda}=2 \epsilon \alpha/(1+\alpha^2)=\epsilon$.
 Treating $\epsilon$ again as a small parameter, we have
\begin{align}
&\Phi(\omega_m, k_\parallel)= \epsilon\pi \left (\omega_0 \gamma\right)^{1/2} T\sum_{m'}\int\frac{d k'_\parallel}{2\pi} \frac{\Phi(\omega_{m'}, k^{'}_{\parallel})}
{ |\omega_{m'} +\Sigma(\omega_{m'}, k^{'}_{\parallel})|} \times \frac{1}{k^2_\parallel + k^{'2}_\parallel +\gamma |\omega_m-\omega_{m^{'}}|}
\label{pairing1}
\end{align}
where the self-energy $\Sigma$
  is given by (\ref{sigma1}).
  The dependence on system parameters $\omega_0$ and $\gamma$ can be eliminated if we measure $\omega$ and $T$ in units of $\omega_0$, and measure $k_\parallel$ in units
  of $(\omega_0 \gamma)^{1/2}$.  Introducing ${\bar \omega} = \omega/\omega_0$, ${{\bar T}} = \pi T/\omega_0$,
   and ${\bar k}_{\parallel} = k_\parallel/(\omega_0 \gamma)^{1/2}$, we re-write (\ref{pairing1}) as
 \begin{align}
&\Phi({\bar \omega}_m, {\bar k}_\parallel)= \epsilon  {{\bar T}} \sum_{m'}\int\frac{d {\bar k}'_\parallel}{2\pi} \frac{\Phi({\bar \omega}_{m'}, {\bar k}^{'}_{\parallel})}
{ |{\bar \omega}_{m'} +{\bar \Sigma}({\bar \omega}_{m'}, {\bar k}^{'}_{\parallel})|} \times \frac{1}{{\bar k}^2_\parallel + {\bar k}^{'2}_\parallel + |{\bar \omega}_m-{\bar \omega}_{m^{'}}|},
\label{pairing_11}
\end{align}
where
\begin{align}
{\bar \Sigma}({\bar \omega}_m,{\bar k}_\|)=\left(\sqrt{|{\bar \omega}|+ {\bar k}^2_\parallel} -|{\bar k}_\parallel| \right)
\label{sigma_1_11}
\end{align}

The question we address is whether $T_c$ is non-zero already at arbitrary small $\epsilon$, or it only emerges when $\epsilon$ exceeds a certain threshold.

Because $\epsilon$ is treated as small parameter, ${{\bar T}}_c$ is expected to be small, and to get the pairing we need to explore logarithmical behavior which comes from frequency scale larger than $T_c$. Accordingly, we replace ${{\bar T}} \sum_{m'}$ by $(1/2) \int d \bar{\omega}_{m'}$ and set $\pm \bar T$ as the lower limits of the integration over positive and negative $\omega_{m'}$, respectively. We then have, instead of (\ref{pairing_11})
 \begin{align}
&\Phi({\bar \omega}_m, {\bar k}_\parallel)= \frac \epsilon 2  \int d{\bar \omega}_{m'} \int\frac{d {\bar k}'_\parallel}{2\pi} \frac{\Phi({\bar \omega}_{m'}, {\bar k}^{'}_{\parallel})}
{ |{\bar \omega}_{m'} +{\bar \Sigma}({\bar \omega}_{m'}, {\bar k}^{'}_{\parallel})|} \times \frac{1}{{\bar k}^2_\parallel + {\bar k}^{'2}_\parallel + |{\bar \omega}_m-{\bar \omega}_{m^{'}}|},
\label{pairing_21}
\end{align}

In general, $\Phi ({\bar \omega}_m, {\bar k}_\parallel)$ is a function of both arguments, but in proper limits the dependence on one of the arguments is stronger than
 on the other. In the main text we consider the limits ${\bar k}^2_\parallel \gg {\bar \omega}_m$ and ${\bar k}^2_\parallel \ll {\bar \omega}_m$.
 In the first case,  the momentum dependence is stronger than frequency dependence, and $\Phi ({\bar \omega}_m, {\bar k}_\parallel)$ can be approximated by $\Phi ({\bar k}_\parallel)$.  The fermionic self-energy at ${\bar k}^2_\parallel \gg {\bar \omega}_m$ behaves as
 \begin{align}
{\bar \Sigma}({\bar \omega}_m,{\bar k}_\|)\approx \frac{|{\bar \omega}_m|}{2|{\bar k}_\parallel|}
\label{sigma_21}
\end{align}
 Substituting this form into (\ref{pairing_21}) we obtain
 \begin{align}
\Phi({\bar k}_\|)= \frac{2 \epsilon}\pi\int \frac{{\bar k}'_\|d{\bar k}'_\|}{{\bar k}_\|^2+{\bar k}_\|^{'2}} \frac{1}{2{\bar k}'_\|+1}\Phi(k'_\|)\int_{{{\bar T}}}^{{\bar k}_\|^2}\frac{d{\bar \omega}^{'}_m}{{\bar \omega}^{'}_m}.
\end{align}
Integrating explicitly over $\omega_m$ and  introducing new variable $x={\bar k}_\|^{'2}/{{\bar T}}$
, we obtain  Eq 6 in the main text.
 \begin{align}
\Phi(y)=\frac\epsilon\pi \int_1\frac{dx}{x+y} \frac{\log{x}}{
2\sqrt{x {{\bar T}}} +1}  \Phi (x),
\label{41}
\end{align}

In the opposite limit, ${\bar k}^2_\parallel \ll {\bar \omega}$, the momentum dependence of the self-energy is small, and ${\bar \Sigma}$ can be approximated by its value at the hot spot, i.e., ${\bar \Sigma} \approx \sqrt{|{\bar \omega}_m|}$.  This limit can only be justified at small or large ratio of $v_y/v_x$ because otherwise
substituting this ${\bar \Sigma}$ into (\ref{pairing_21}) we obtain that typical internal ${\bar k}^{'2}_{\parallel}$ are of order ${\bar \omega}^{'}$.
Nevertheless, if we assume that the momentum dependence of the fermionic self-energy can be neglected, at least for order-of-magnitude estimates, we obtain that
 $\Phi ({\bar \omega}_m, {\bar k}_\parallel)$ can be approximated by $\Phi ({\bar \omega}_m)$, and the gap equation becomes~\cite{acs1,acf1}
 \begin{align}
\Phi({\bar \omega}_m)= \frac {\pi \epsilon {{\bar T}}}2 \sum_{m'}\frac{\Phi({\bar \omega}_{m'})}{\sqrt{|{\bar \omega}_{m'}|}\sqrt{|{\bar \omega}_{m}-{\bar \omega}_{m'}|} (1 + \sqrt{|{\bar \omega}_{m'}|})}.
\end{align}
This is Eq. (9) in the main text.

\subsection{Solution of the Eq. \ref{41}}

We first replace the soft cutoff imposed by $\left(
2\sqrt{x{\bar T}}+1\right)^{-1}$ with a hard cutoff,
\begin{align}
\Phi(y)=\frac\epsilon \pi \int_1^{1 /{{\bar T}}}\frac{dx}{x+y}\log x~\Phi(x)
\label{ch_71}
\end{align}
The integration over $x \gg y$ gives rise to $\log^2$ terms in the perturbation theory, which, as we know from the analysis in the previous section, do not give rise to
 the pairing instability. We therefore focus on the range $ x \sim y$. The contribution from this range gives rise to a single logarithm ($\log 1/{{\bar T}}$) if we
  momentary assume that $\Phi (x)$ is a constant.  Like we did before, we treat $\epsilon$ as a small parameter and check whether Eq. (\ref{ch_71}) has a solution at a finite ${{\bar T}}$.

For small $\epsilon$, ${\bar T}$ is expected to be also small, i.e., the upper limit of the integration in (\ref{ch_71}) is a large number.
 We first consider $y$ and $x \sim y$ in the range  $1\ll y,x \ll 1/{\bar T}$, and search for the solution of (\ref{ch_71}) in the form
 $\Phi(x)= \exp[-f(p(x))]$, where $p(x) = \epsilon \log{x}$. Plugging this function back into the equation and introducing a new variable $z=x/y$,
 we obtain,
\begin{align}
e^{-f( \epsilon\log y)}=\frac\epsilon\pi\log y\int_{\frac 1 y}^{\infty}\frac{dz}{z+1}\left(1+\frac{\log z}{\log y}\right)e^{-f( \epsilon (\log y+\log z))}
\end{align}
We assume and then verify that  typical $z$ are of order one, i.e., $\log z \ll \log y$.
Taylor expanding in $\log z/\log y$ we obtain
\begin{align}
1=&\frac\epsilon\pi\log y~\int_0^\infty\frac{dz}{z+1}\frac{1}{z^{Q(\epsilon\log y)}}\nonumber\\
=&\epsilon\log y~\frac{1}{\sin(\pi  Q(\epsilon \log y))}.
\label{ch_81}
\end{align}
where we introduced
\begin{align}
Q(p)\equiv \frac{df(p)}{dp}.
\label{nn_11}
\end{align}
Solving (\ref{ch_81}) we obtain
\begin{align}
Q(\epsilon \log y)=\frac{1}{\pi}\arcsin(\epsilon\log y).
\label{ch_101}
\end{align}
Integrating over $\log y$, we have
\begin{align}
f(p)=\frac{1}{\pi\epsilon}\left[p\arcsin\left(p\right)+\sqrt{1-p^2}-1\right],
\label{ch_91}
\end{align}
which is Eq 7 of the main text.

Lastly we consider the boundary condition at $y=y_0=1/{{\bar T}}$. For such $y$, the upper limit is 1.
Substituting formally the trial solution  $\Phi(x)= \exp[-f(p(x))]$  into
 the r.h.s. of (\ref{ch_71}) we obtain
\begin{align}
1=\frac\epsilon\pi\log y_0~\int_0^1\frac{dz}{z+1}\frac{1}{z^{Q(2\pi \epsilon \log y_0)}}
\label{ch_111}
\end{align}
One can easily make sure that to satisfy this equation, $Q(\epsilon \log y_0)$ must be
 {\it larger} than
 if we take our result, Eq. (\ref{ch_101}), and  set $y = y_0$ there.
 At vanishingly small  ${\bar T}$, the presence of the upper limit of the integration over $x$ in (\ref{ch_71}) becomes relevant only
  for $y$ infinitesimally close to $y_0$ (for smaller $y$, the upper limit can be safely set to infinity).
  This implies that at ${\bar T} \to 0$,  $Q(p)$  undergoes a finite jump at $y=y_0$, hence at the actual $T=T_c$
\begin{align}
\frac{dQ(p)}{dp}\bigg|_{p=\epsilon \log{1/\bar T}_c}=\infty.
\end{align}
 This last equation is satisfied if $\epsilon\log 1/{\bar T_c}=1$
  Restoring the parameters, we then obtain  Eq 8 in the main text.
 For this $T_c$, the jump in $Q$ at the boundary is between $Q =0.5$ (Eq. (\ref{ch_101}) at $y =y_0 = 1/{\bar T}_c$) and $Q =0.73$, which is the solution of
 (\ref{ch_111}) at $T_c$.

  \begin{figure}[h]
\includegraphics[width=.5\columnwidth]{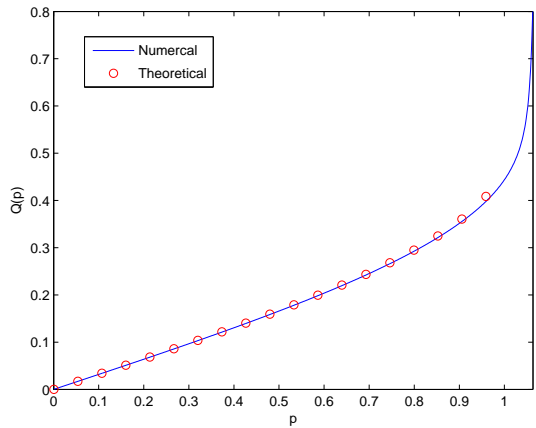}
\caption{Numerical solution of $Q(p)$ and its comparison with theory. We  set  $\epsilon=1.5\times10^{-2}$.}
\label{fig:s21}
\end{figure}

 In Fig. \ref{fig:s21} we plot $Q(p)$ obtained from the numerical solution of Eq. (\ref{ch_71}) and compare it with the (approximate) analytical solution presented in this section. We see that the agreement is quite good.

 \subsection {Gap equation for $\alpha\neq 1$} \label{neq11}

The reasoning we used in the previous section can be
 extended to the case of a general $\alpha$.

We follow the same procedure leading to Eq. \ref{ch_71}, only this time we keep explicit $\alpha$ dependence along the way. We introduce new, $\alpha$-dependent  variable $x$ as $x=2\alpha/(\alpha^2+1)~\bar{k}_\|^{'2}/\bar T$ and obtain an $\alpha$ dependent version of Eq. \ref{ch_71},
\begin{align}
\Phi(y)=\frac{\bar\lambda}\pi\int_1^{1/\bar T}\frac{(x+y)~dx}{(x+y)^2-4\mu^2xy}\log x~\Phi(x),
\label{genal1}
\end{align}
where $\mu=(1-\alpha^2)/(1+\alpha^2)$ and, we remind, ${\bar \lambda}=2\epsilon \alpha/(1+\alpha^2)$. We again assume and then verify  that the major contribution  to the r.h.s. of (\ref{genal1}) comes from $x\sim y$, i.e., $z\sim 1$, and, as before, introduce $p = \epsilon \log y$ and the same $Q(p)$ as in (\ref{nn_11}).  We then obtain
 for $1\ll y,x \ll 1/\bar T$,
\begin{align}
1=\frac{\epsilon \lambda}{\pi}\log y\int_0^\infty \frac{(z+1)~dz}{(z+1)^2-4\mu^2z}\frac{1}{z^{Q(\epsilon\log y)}}.
\end{align}
or
\begin{align}
p=\frac\pi {\lambda I(Q(p),\mu)},
\label{pI1}
\end{align}
where
\begin{align}
I(Q,\mu) =\int_0^\infty \frac{(z+1)~dz}{(z+1)^2-4\mu^2z}\frac{1}{z^{Q(\epsilon\log y)}}.
\label{nn_21}
\end{align}
We know from previous section that the upper
   boundary for $x \sim y$ is $x,y \sim 1/{\bar T}_c$, or $p= \epsilon \log{1/{\bar T}_c}$, and that at the upper boundary
  $\frac{dQ(p)}{dp}\big|_{p=\epsilon \log{1/\bar T}_c}=\infty$, or, $\frac{dp}{dQ}\big|_{p=\epsilon \log{1/\bar T}_c}=0$. Combining this with
   Eq. \ref{pI1}, we obtain
\begin{align}
\frac{dp}{dQ}\bigg|_{p=\epsilon \log{1/\bar T}_c}\propto\frac{dI}{dQ}\bigg|_{p=\epsilon \log{1/\bar T}_c} {=}0.
\end{align}
  Hence, at $y=\bar{T}_c$, ${dI}/{dQ}=0$.
  It is easy to see from (\ref{nn_21})
  that $I(Q)=I(1-Q)$, therefore
   the boundary condition sets
   $Q=1/2$. Plugging $p=\epsilon \log{1/\bar T}_c$ and $Q=1/2$ back to Eq. \ref{pI1}, we obtain
\begin{align}
\epsilon \log{1/\bar T}_c=\frac\pi {\lambda I(Q=1/2,\mu)}.
\label{nn_31}
\end{align}
 The integral for $I(Q=1/2,\mu)$ can be easily evaluated, and we obtain
\begin{align}
 I(Q=1/2,\mu)&=\int_0^\infty \frac{1+z}{(1+z)^2-4\mu^2z}~\frac{dz}{z^{1/2}} \nonumber \\
 &=\frac{\pi}{\sqrt{1-\mu^2}}\nonumber \\
 &=\frac{\pi}{\lambda}.
\end{align}
 where in the last step we used the fact that $\mu^2+\lambda^2=1$. Substituting this finally into (\ref{nn_31}) we obtain
 $\epsilon\log 1/\bar T_c=1$, i.e., $T_c \sim {\omega_0} e^{-1/\epsilon}$, independent on $\alpha$.

 We emphasize that independence of $T_c$ on  $\alpha$  could not be seen from naive perturbation analysis, since back there the prefactor
 ${\bar \lambda}$ for the $\log^2$ term was $\alpha$ dependent. For verification, we solved Eq \ref{genal1} numerically for various $\alpha$
  and indeed find that $T_c$ does not change when $\alpha$ changes.

\subsection{The case of strong nesting}

\begin{figure}[htbp]
\includegraphics[width=.5\columnwidth]{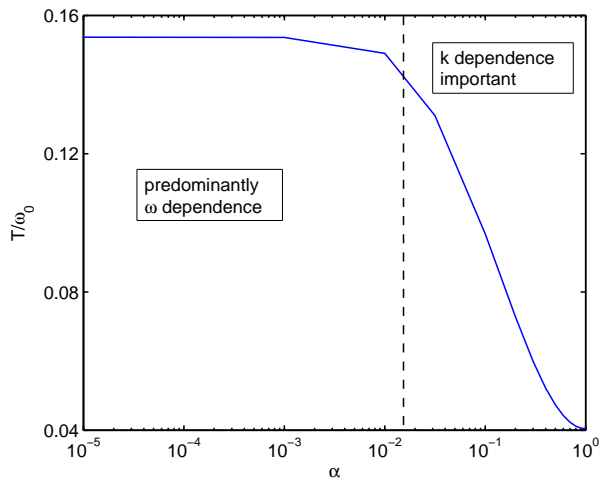}
\caption{Numerical solution of Eq. (\ref{pairing1}) for various
 $\alpha= v_y/v_x$ (the Fermi velocities at hot spots separated by $(\pi,\pi)$ are ${\bf v}_{F,1} = (v_x, v_y)$ and
   ${\bf v}_{F,2} = (-v_x, v_y)$.
    We set ${\bar g} a/v_F\approx0.032$.
For
$\alpha = O(1)$ the momentum dependence of the fermionic self-energy is relevant, and $T_c\approx 0.04 \omega_0$.
At smaller $\alpha$, when $\alpha <  ({\bar g}/(v_F/a))^{2/3}$ (to the left of vertical dashed line), frequency dependence of the self-energy prevails, the pairing comes exclusively from fermions with non-FL self-energy, and
$T_c \approx 0.17 \omega_0$, up to corrections of order $({\bar g} a/v_F)^{1/2}$. In our case, these corrections
 reduce this number
 from $0.17$ to $0.15$.}
\label{fig:contour1}
\end{figure}

At strong nesting, when velocities at ${\bf k}_F$ and ${\bf k}_F + {\bf Q}$ are nearly antiparallel,
$\alpha \ll1$ and $\gamma \propto 1/\alpha$ is large. In this situation, the pairing eventually become determined by
fermions whose  self-energy has a  non-FL form.
 This happens when $(\gamma \omega_0)^{1/2}$ becomes larger than maximal possible
 $\alpha|k_\parallel|$ along a Fermi arc, which is of order $1/a$, i.e., when
 $v_F/a \gg {\bar g} > \alpha v_F/a $.  In this situation, the momentum dependence of $\Sigma (\omega_m, k_\parallel)$ becomes irrelevant at $\omega_m \sim \omega_0$, $\Phi (\omega_m, k_\parallel)$ becomes predominantly frequency-dependent, and $T_c$ recovers the  value $0.17 \omega_0$, which is $T_c$ for momentum-independent self-energy.
 We verified numerically that this is indeed the case. We plot
   $T_c$ vs $\alpha$ in Fig. \ref{fig:contour1}.
   At $\alpha = O(1)$, $T_c \approx 0.04 \omega_0$, and at small enough $\alpha$, $T_c$ approaches $0.17 \omega_0$.
     We caution, however, that  the limit $\alpha \ll1$ has to be taken with care as nesting may generate additional singularities at higher-loop orders.

\end{document}